\documentclass[fleqn,twoside]{article}
\usepackage{espcrc2}

\usepackage{graphicx}
\usepackage[figuresright]{rotating}
\def\be{\begin{eqnarray} &&}
\def\nonu{\nonumber \\ &&}
\def\ee{\end{eqnarray}}
\newcommand{\AmS}{{\protect\the\textfont2
  A\kern-.1667em\lower.5ex\hbox{M}\kern-.125emS}}

\title{Light-front quark distributions in the nucleon  
and nucleon electromagnetic  form factors}

\author{E. Pace \address{Dipartimento di Fisica, Universit\`a degli Studi di Roma "Tor Vergata" and Istituto
Nazionale di Fisica Nucleare, Sezione Tor Vergata, Via della Ricerca
Scientifica 1, 00133  Roma, Italy},
        J. P. B. C. de Melo \address{Centro de Ci\^encias Exatas e Tecnol\'ogicas,
 Universidade Cruzeiro do Sul, 08060-070, S\~ao Paulo, Brazil},
        T. Frederico \address{Dep. de F\'\i sica, Instituto Tecnol\'ogico de Aeron\'autica,
 12.228-900 S\~ao Jos\'e dos
Campos, S\~ao Paulo, Brazil},
S. Pisano \address{IN2P3, Institut de Physique Nucl\'eaire d'Orsay, 91404 Orsay, France}
        and
        G. Salm\`e \address{Istituto  Nazionale di Fisica Nucleare, Sezione di Roma, P.le A. Moro 2,
 00185 Roma, Italy}}
       
\begin{document}

\begin{abstract}
 Longitudinal and transverse quark momentum distributions in the nucleon 
 are calculated from a phenomenological quark-nucleon vertex
function obtained through an investigation of the nucleon electromagnetic form factors within 
a light-front framework.
\vspace{1pc}
\end{abstract}

\maketitle

\section{INTRODUCTION}

A wealth of information on the partonic structure of the nucleon is encoded in the generalized parton
distributions (GPD's) and extensive research programs are being pursued to gain information on nucleon 
GPD's. 

Our strategy to this end is to determine the quark-nucleon vertex function from an investigation of nucleon
electromagnetic (em) form factors within the light-front dynamics, and then to use the obtained vertex
function to evaluate the nucleon GPD's. Indeed, light-front dynamics opens a unique possibility to study
the hadronic state in both the valence and the nonvalence sector \cite{Bro}, since within a light-front
framework no spontaneous pair production occurs and a meaningful Fock state
expansion is possible :
\be
| baryon \rangle =  {|qqq \rangle} + 
{|qqq~q \bar{q} \rangle +|qqq~g \rangle
 .....}
 \label{Fock}
\ee
As a first step, in this contribution we present our preliminary results for the unpolarized longitudinal
and transverse parton momentum distributions in the nucleon.

\section{NUCLEON VERTEX FUNCTION AND NUCLEON ELECTROMAGNETIC FORM FACTORS}
We describe the quark-nucleon vertex function through a Bethe-Salpeter amplitude (BSA),
whose Dirac structure is suggested
by an effective Lagrangian \cite{de}.
 Then the symmetrized BSA for  the nucleon is approximated as follows
\be
\hspace{-0.6cm}\Phi^\sigma_N(k_1,k_2,k_3,P) 
=\nonu 
\imath 
\left [~S(k_1)~\tau_y ~ \gamma^5 ~ S_C(k_2)C ~\otimes~ S(k_3)~+
~\right. \nonu 
\hspace{-0.0cm}S(k_3)~ \tau_y ~ \gamma^5 ~S_C(k_1)C ~\otimes~ S(k_2)~+
\nonu 
 \left.~S(k_3)~ \tau_y ~ \gamma^5 ~S_C(k_2)C~\otimes~S(k_1)
 \right ]  \nonu
\times ~~ \Lambda(k_1,k_2,k_3) ~ \chi_{\tau_N} ~ U_N(P,\sigma)
\ee
where $S(k)$ is the constituent quark (CQ) free propagator,
$S_C(k)$ the charge conjugated propagator,
$\Lambda(k_1,k_2,k_3)$ describes  the symmetric momentum dependence of the
vertex function upon the quark momentum variables, $k_i$,
$U_N(P,\sigma)$ is the nucleon spinor and $\chi_{\tau_N}$ the isospin eigenstate.

 The matrix elements of the {\em{macroscopic}}  current in the spacelike (SL) region
 are approximated {\em{microscopically}} by the Mandelstam formula \cite{mandel}
\vspace{0.0cm}
\be
\hspace{-0.7 cm}\langle \sigma',P'|j^\mu~|P,\sigma\rangle
=~3~N_c ~ \times 
\nonu \int {d^4k_1 \over (2\pi)^4}\int {d^4k_2 \over (2\pi)^4} ~ 
{\Large{\Sigma}} \left \{ \bar \Phi^{\sigma'}_N(k_1,k_2,k'_3,P')
 \right. ~ \times
 \nonu \hspace{-0.7 cm}  \left. 
 S^{-1}(k_1) ~ S^{-1}(k_2)~{\cal I}^\mu(k_3,q)~
 \Phi^\sigma_N(k_1,k_2,k_3,P)\right \}
\ee
where $N_c$ is the number of colors, $k'_{3} = k_{3} + q$, and ${\cal I}^\mu(k_3,q)$ 
is the quark-photon vertex. 
An analogous expression holds
 in the timelike (TL) region.
 
We adopt a Breit reference frame where~$~~{\bf q}_{\perp}=0$ and
$q^+=q^0 + q^3=|q^2|^{1/2}$. Our CQ mass is ~$m=m_u = m_d = 200 ~ MeV$.

The 
Mandelstam Formula is projected out by an analytic integration on $k_1^-$ and  $k_2^-$,
 taking into account only the poles of the propagators. 
Then the current becomes the sum of a purely valence contribution (diagram (a) in Fig. 1)
and a nonvalence (NV), pair-production contribution (diagram (b) in Fig. 1).
Clearly, after the $k^-$ integrations,
the vertex functions depend only upon the light-front three-momentum.
\begin{figure}

\includegraphics[width=7.5cm]{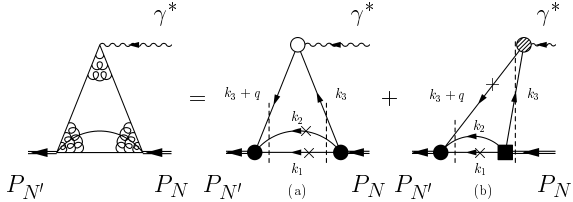}

\vspace{-0.6cm}

\caption{Diagrams for the SL nucleon current: (a) valence (triangle) contribution with 
$0 < k_{i}^+ < P^+$, and $~0 \le k_{3}^+ + q^+ < {P'}^{+}$;
 (b) nonvalence, pair-production contribution with $0 > k_{3}^+ \ge -q^+$. 
 A cross indicates a quark on the $k^-$-shell, i.e.  $k^- = k^-_{on} = (m^2 + k^2_{\perp})/k^+$.
 Solid circles and solid square represent valence and NV vertex functions, respectively
 (after Ref. \cite{nucleon}). } 
\end{figure}

The quark-photon vertex has isoscalar and isovector contributions
\be
  {\cal I}^\mu=~{\cal I}^\mu_{IS} +\tau_z
  {\cal I}^\mu_{IV}
  \label{curr}
 \ee
and each term in Eq.  (\ref{curr}) contains a purely valence contribution 
  (in the SL region only) and a contribution corresponding to the pair production (Z-diagram). 
  In turn the Z-diagram contribution can be decomposed in a bare term $+$ a 
  Vector
  Meson Dominance (VMD) term (according to  the decomposition of the photon
  state in bare,  hadronic  [and leptonic] contributions), viz
\be
\hspace{-0.7 cm} {\cal I}^\mu_{i}(k,q) = {\cal N}_{i} ~ \theta(P^+-k^+)\theta(k^+)
   \gamma^\mu+\theta({q}^+ + k^+)
  \nonu\hspace{-0.7 cm} \times ~
 \theta(-k^+)~\left \{ Z_B~{\cal N}_{i} ~ \gamma^\mu+ 
  Z^i_{VM}~\Gamma^\mu[k,q,i]\right\} 
   \ee
 with  $i = IS, IV$, ${\cal N}_{IS}=1/6$ and ${\cal N}_{IV}=1/2$.   The constants
 $Z_B$ (bare term) and $Z^i_{VM}$ (VMD term) are unknown weights 
 to be extracted from
 the phenomenological analysis of the data.

 According to $i$, the VMD term $\Gamma^\mu[k,q,i]$ 
  includes isovector or isoscalar mesons. Indeed in \cite{nucleon} we extended to isoscalar mesons
   the microscopic model for the VMD successfully used in
    \cite{DFPS} for the pion form factor in the SL and in the TL region
  and based on the meson mass operator of Ref. \cite{Fred}.
   As explained in \cite{nucleon}, $\Gamma^\mu[k,q,i]$
 does not involve free parameters. We consider up to 20 mesons for achieving convergence at high $|q^2|$.
  
  In the valence vertexes (solid circles in Fig. 1)  the spectator quarks are on the
$k^-$-shell, and the BSA
momentum dependence  
is approximated through a
nucleon wave function a la Brodsky (PQCD inspired), namely
\be
\hspace{-0.6cm} \Psi_N(\tilde{k}_1,\tilde{k}_2,P)  = P^+ ~
{ {\Lambda}(k_1,k_2,k_3)|_{(k_{1on}^-,k_{2on}^-)} \over [m_N^2 - M^2_0(1,2,3)]^{~}} = 
\nonu 
\nonu  = ~  P^+ ~ {\cal {N}}~
{~(9~m^2)^{7/2}
 \over (\xi_1\xi_2\xi_3)^{p}~ \left[\beta^2 + M^2_0(1,2,3)\right]^{7/2}}~
\ee
where $\tilde{k}_i \equiv (k_i^+,{\bf k}_{i\perp})$, $M_0(1,2,3)$ 
is the free mass of the three-quark system,  
$\xi_i = k^+_i/P^+$ \quad ($i=1,2,3$)
and  ${\cal N}$  a normalization constant.

The power $ 7/2 $ and the parameter $ p = 0.13 $ are chosen to have
an asymptotic decrease of the triangle contribution faster than the dipole.

Only the triangle diagram determines
the magnetic moments, which are weakly dependent on $p$. Then
$\beta = 0.645$ can be fixed by the magnetic moments and we obtain $\mu_p = 2.87\pm0.02$ ~
($\mu_p^{exp}$ = 2.793)
and $\mu_n = -1.85\pm0.02$ ~ ($\mu_n^{exp}$ = -1.913).
\begin{figure}[htb]
\vspace{-0.0cm}
\hspace{-.0cm}{\includegraphics[width=7.5cm]{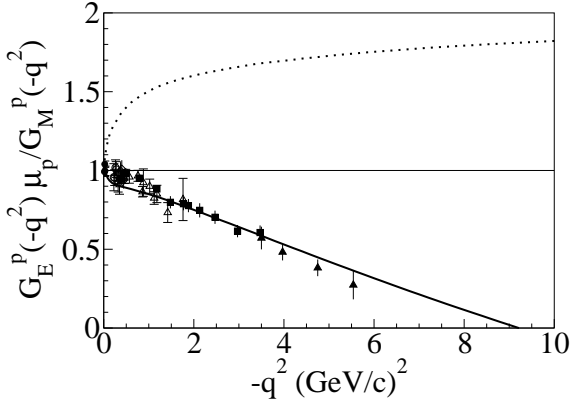}}
\vspace{-1.3cm}
\caption{Ratio between electric and magnetic form factors for  
the proton vs $-q^2$. Solid line: full calculation, sum of triangle and 
pair-production terms;  dotted line: triangle contribution only (after Ref. \cite{nucleon}). } 
\end{figure}
\begin{figure}[htb]
\hspace{+.0cm} 
{\includegraphics[width=7.5cm]{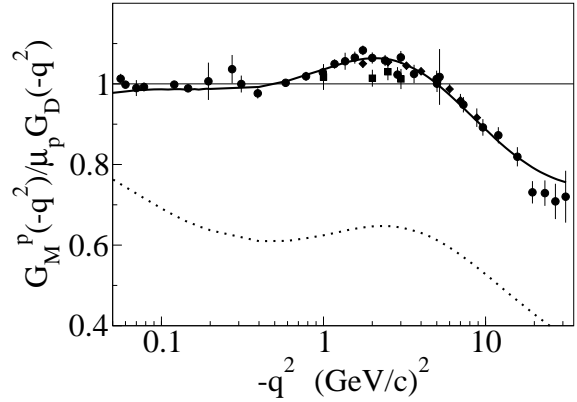} }
\vspace{-1.3cm}
\caption{Magnetic  proton form factor vs $-q^2$. Solid and dotted lines  as 
in Fig. 2. $G_D(|q^2|) = [1 + |q^2|/(0.71$ (GeV/c)$^2]^{-2}$  (after Ref. \cite{nucleon}). } 
\end{figure}

For the Z-diagram contribution, the NV vertex (solid square in Fig. 1)
is needed.
 It can depend on the available invariants, i.e.
  on the free mass, $M_0(1,2)$, of the (1,2) quark pair 
and on the free mass, $M_0(N,\bar {3})$, of the ( nucleon - quark $\bar {3}$ ) system 
entering the NV vertex.
Then in the SL region we approximate the
 momentum dependence of the NV vertex 
 $ {\Lambda}_{NV}^{SL} = {\Lambda}(k_1,k_2,k_3)|_{k^-_{1on},k^{'-}_{3on}}$ by
\begin{eqnarray}
&&\hspace{-.9cm} {\Lambda}_{NV}^{SL}= [g_{12}]^{2}[g_{N\bar {3}}]^{3/2}
 \left [{k_{12}^+ \over P^{\prime +} }  \right ]
  \left [ P^{\prime +} \over k_{\overline {3}}^+ \right ]^r
  \left [P^{+} \over  k_{\overline {3}}^+   \right ]^{r}  
\end{eqnarray}
with
\begin{eqnarray}
 k_{12}^+ = k_1^+ +  k_2^+ ~~ \quad  g_{AB} = {(m_A ~ m_B) \over \left
[\beta^2+M^2_0(A,B)\right]}
\end{eqnarray}
The power 2 of $[g_{12}]^{2}$ is suggested from counting rules. 
The power 3/2 of $[g_{N\bar {3}}]^{3/2}$ and the parameter $r=0.17$ are chosen
 to have an asymptotic dipole behavior for the NV contribution.
\begin{figure}[htb]
\vspace{-0.0cm}
\hspace{-.0cm}
{\includegraphics[width=7.cm]{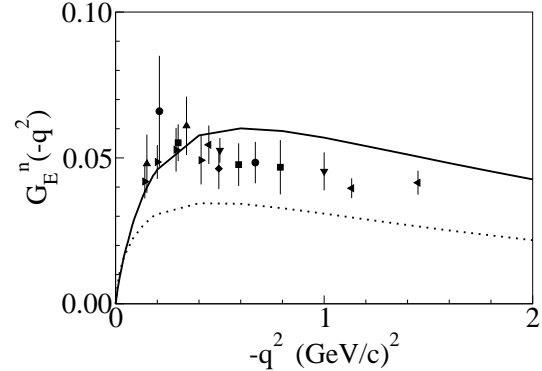}}
\vspace{-0.9cm}
\caption{Electric neutron form factor vs $-q^2$.  
Solid and dotted lines as in Fig. 2 (after Ref. \cite{nucleon}). } 
\end{figure}
\begin{figure}[htb]
\hspace{+.0cm} 
{\includegraphics[width=7.cm]{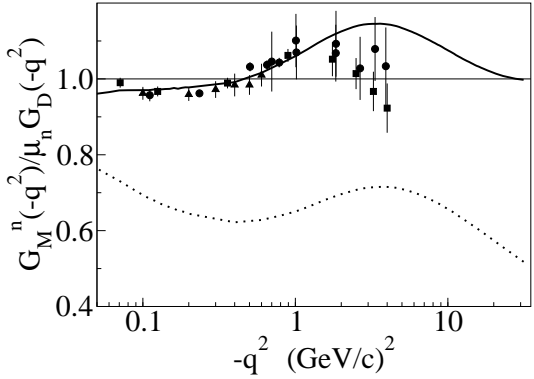} }
\vspace{-0.9cm}
\caption{The same as in Fig. 3, but for the neutron (after Ref. \cite{nucleon}). } 
\end{figure}

 Analogous expressions with the same parameters are used for the nonvalence vertexes in the TL region 
 (see Ref. \cite{nucleon}
 for the explicit expressions). 
 
 We perform a fit of our free parameters, $Z_B$, $Z^i_{VM}$, $p$, $r$  in the SL region 
 and obtain the form factors shown in Figs. 2 - 5,
 with a $\chi ^2$/datum = 1.7. As a result
 the weights for the pair-production terms 
are $Z_B = Z_{VM}^{IV} = 2.283$  and 
$Z_{VM}^{IS} / Z_{VM}^{IV} = 1.12$, remarkably close to one.
The same values of our weight parameters are adopted to evaluate the TL form factors shown in Figs. 6 and 7.

Preliminary results of our model for the nucleon form factors were presented in \cite{MFPPS}.

The Z-diagram turns out to be essential for the description of the form factors in our reference frame
with $q^+ > 0$. 
In particular the possible zero in $G_E^p/G_M^p$ is strongly related to the pair-production contribution.

In the TL region our parameter free calculations give a fair description of the proton data,
apart for some missing strength at $q^2 = 4.5$ (GeV/c)$^2$ ~ and ~ $q^2 = 8$ (GeV/c)$^2$ (as 
occurs for the pion
case \cite{DFPS}), which one could argue to be due to possible unknown vector mesons, missing
in the spectrum of Ref. \cite{Fred}.
\begin{figure}[htb]
\centering
\includegraphics[width=7.cm]{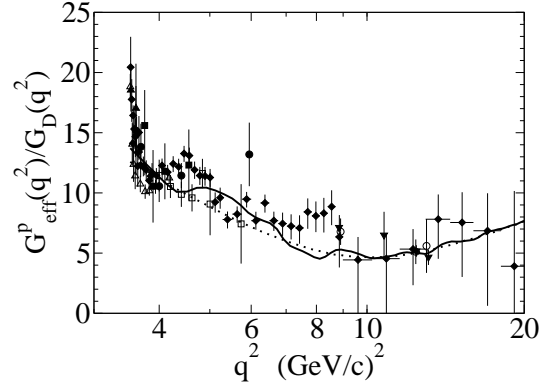} 
\vspace{-.9cm}
\caption{Proton  effective form factor $G^p_{eff}(q^2)/G_D(q^2)$ in the timelike region vs $q^2$. 
Solid line: bare +
VMD; dotted line: bare term.
$G_{eff}(q^2) = [(|G_M(q^2)|^2 ~ - ~ \eta ~|G_E(q^2)|^2  ) / 
(1 ~ - ~ \eta)]^{0.5}$ with $\eta = - 2 m_N^2 / q^2$ (after Ref. \cite{nucleon}). } 
\end{figure}
\begin{figure}[htb]
\includegraphics[width=7.cm]{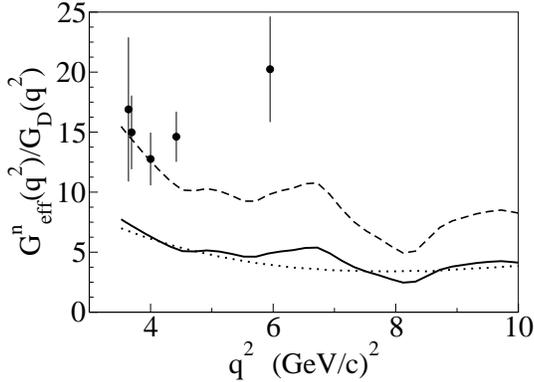}
\vspace{-.9cm}
\caption{The same as in Fig. 5, but for 
the neutron. Dashed line: solid line arbitrarily multiplied by 2 (after Ref. \cite{nucleon}). } 
\end{figure}

\section{LONGITUDINAL AND TRANSVERSE QUARK MOMENTUM DISTRIBUTIONS IN THE NUCLEON}
 The longitudinal distribution $q(x)$ is the limit in
     the forward case, $P' = P$,  of the unpolarized 
     generalized parton distribution ${H}^q(x,\xi,t)$.
  Indeed one can define the distributions ${H}^q(x,\xi,t)$ and $E^q(x,\xi,t)$ for the quark $q$
  through the following relation  
    \begin{eqnarray}
   \label{gpdisos}
 \frac{1}{2P^+}  \bar{u}(P',\lambda')\left [ {H}^q~ \gamma^+  + 
 {E}^q ~\imath  \frac{\sigma^{+\alpha}q_{\alpha}}{2 M} \right ] {u}(P,\lambda)=  \nonumber
 \\   \int \frac{dz^-}{4\pi} e^{i x {\cal{P}}^+ z^-} 
 \left . \langle P',\lambda' | \bar \psi_q
(-\frac{z}{2} ) \gamma ^+ \,  \psi_q (\frac{z}{2} ) | P,\lambda
\rangle \right |_{\tilde{z}=0} \nonumber
\end{eqnarray}
where  $\lambda$ is the nucleon helicity,
$\tilde z\equiv \{z^+, {\bf z}_{\perp}\}$ , 
$\psi_q(z)$ is the quark field  isodoublet, while
   \be
   {\cal{P}}=\frac12(P'+P)
  \;\; \quad \quad t=(P' - P)^2 
  \label{kin1}
  \ee
  \be
 \xi=-\frac{P'^+ - P^+}{2 {\cal{P}}^+} \quad \quad  
x=\frac{k^+}{{\cal{P}}^+}
\label{kin2}
\ee
and $k$ is the average
momentum of the quark that interacts with the photon, i.e.
  \be 
  k = \frac{k_3 + (k_3 +q)}{2}
  \ee
For $P' = P$, both $q^+$ and $\xi$ are vanishing and $x = k_3^+ / P^+ = \xi_3$
coincides with the fraction of the longitudinal momentum carried by the active quark, i.e., with the
Bjorken variable. As a consequence the function ${H}^q(x,\xi,t)$ reduces to the longitudinal parton
distribution function $q(x)$ :
\begin{eqnarray}
\hspace{-2cm}  {H}^q(x,0,0) = q(x) = \int d{\bf k}_{\perp} ~~ f_1^q(x,k_{\perp}) =&&
 \label{struc}    \\ 
  \int \frac{dz^-}{4\pi} e^{i x P^+ z^-} 
 \left . \langle P | \bar \psi_q
(-{z\over 2}) \gamma ^+ \,  \psi_q ( {z\over 2}) | P \rangle \right |_{\tilde{z}=0} \nonumber
  \end{eqnarray}
    where 
    an average on the nucleon helicities is understood.
   
Once all the parameters of the nucleon light-front wave function 
   $\Psi_N(\tilde k_1,\tilde k_2,P)$ have been determined, one can easily define
 the transverse-momentum-dependent distributions 
  of the active quark
  in terms of the nucleon light-front wave function :
 \be
\hspace{-0.6cm} f_1^u(x,k_{\perp})=-{9~N_c\over 32~(2 \pi)^6}~{1\over P^{+2}} ~ \times ~\nonu 
\hspace{-0.7cm}\int^{1 - x}_0   d\xi_2 {1 \over (1 - x -\xi_2) \xi_2} {1 \over 
x^2}
\int d{\bf k}_{2 \perp} \left. {\cal H}_{u}
\right|_{(k^-_{1on},k^-_{2on})}
   \nonu 
   \nonu 
\hspace{-0.3cm} \times ~ |\Psi_N(P^+ ~ \xi_1,{\bf
  k}_{1\perp},P^+ ~ \xi_2,{\bf k}_{2\perp},P)|^2
\label{partutr}
\ee
\be
\hspace{-0.6cm} f_1^d(x,k_{\perp})={9~N_c\over 8~(2 \pi)^6}~ {1\over P^{+2}} ~ \times ~\nonu 
\hspace{-0.7cm}\int^{1 - x}_0   d\xi_2 {1 \over (1 - x-\xi_2) \xi_2} {1 \over 
x^2}
\int d{\bf k}_{2 \perp} \left. {\cal H}_{d}
\right|_{(k^-_{1on},k^-_{2on})} 
   \nonu
   \nonu 
\hspace{-0.3cm}\times ~ |\Psi_N(P^+ ~ \xi_1,{\bf
  k}_{1\perp},P^+ ~ \xi_2,{\bf k}_{2\perp},P)|^2~~
     \quad
\label{partdtr}
\ee  
where  ${\cal H}_{u}$ and ${\cal H}_{d}$ are 
proper traces of propagators and of the 
currents ${\cal {I}}^+_u$ and ${\cal {I}}^+_d$, respectively. 

From the nucleon light-front wave function 
   $\Psi_N(\tilde k_1,\tilde k_2,P)$ one can easily define through Eq. (\ref{struc})
 also  the longitudinal  distribution of the stuck quark and from the isospin symmetry one has
\be
\hspace{-0.7cm}u_p(x)=d_n(x)=u(x); ~
d_p(x)=u_n(x)= d(x)
\ee
Our preliminary results for $f_1^{u(d)}(x,k_{\perp})$ in the proton and for $u(x)$ and $d(x)$  are
shown in Figs. 8, 9  and in Figs. 10, 11, respectively.
\begin{figure}[htb]
\vspace{-.8cm}
\hspace{-.0cm}{\includegraphics[width=7.5cm]{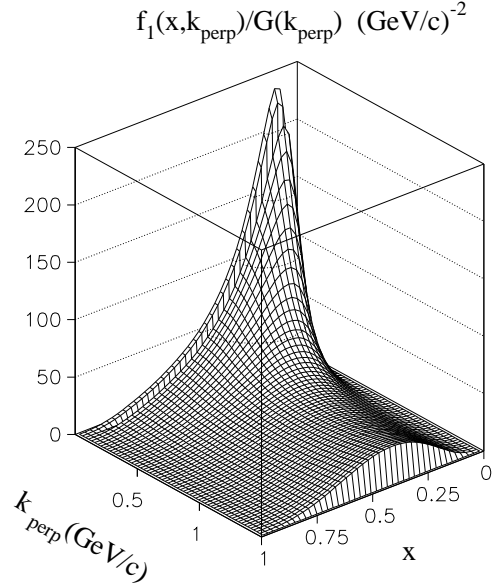}}
\vspace{-2.5cm}
\caption{Valence transverse-momentum distributions for a $u$ quark inside  
the proton. $G(k_{\perp}) = (1 ~ + ~ k_{\perp}^2/m_{\rho}^2)^{-5.5}$,
$m_{\rho}$ = 770 MeV and $k_{perp} = | k_{\perp} |$. } 
\end{figure}
\begin{figure}[htb]
\vspace{-.8cm}
{\includegraphics[width=7.5cm]{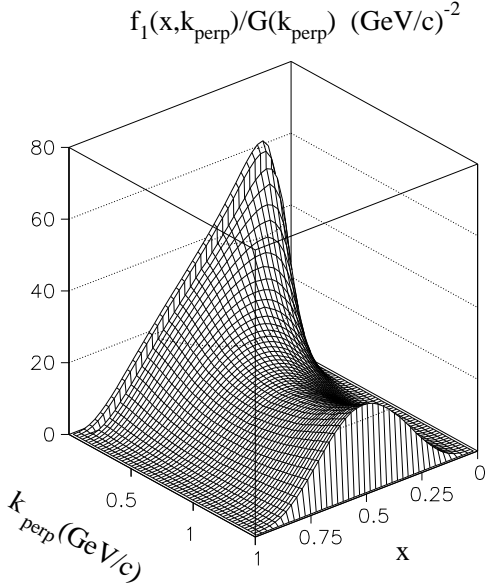}}
\vspace{-2.5cm}
\caption{The same as in Fig. 8, but for a $d$ quark inside  
the proton. } 
\end{figure}
It can be observed that the decay of our $f_1(x,k_{\perp})$ vs $k_{\perp}$ 
is faster than in diquark models of nucleon
\cite{Jacob}, while it is slower than in factorization
models for the transverse momentum distributions \cite{Anselm}.

As far as the longitudinal momentum distributions are concerned, 
a reasonable agreement of our $u(x)$  with the CTEQ4 fit to the experimental data \cite{Lai} 
can be seen in Fig. 10.

\begin{figure}[htb]
\centering
\hspace{.0cm}{\includegraphics[width=7.cm]{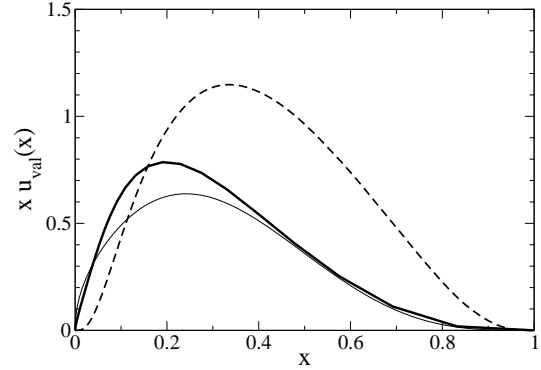}}
\vspace{-.9cm}
\caption{Longitudinal momentum distribution for a $u$ quark inside  
the proton. Dashed lines: our preliminary results; thick solid lines: our results after
evolution to $Q^2$ = 1.6 (GeV/c)$^2$; thin solid lines: CTEQ4 fit to the experimental data \cite{Lai}.} 
\end{figure}
\begin{figure}[htb]
\hspace{.0cm}{\includegraphics[width=7.cm]{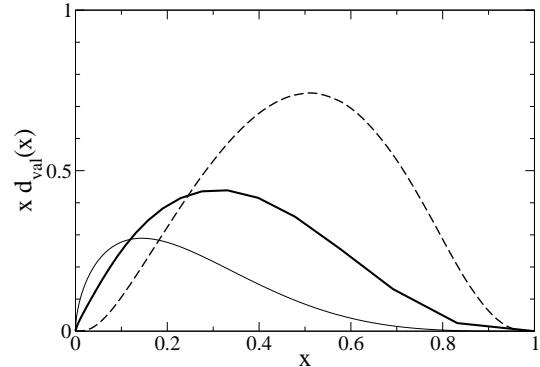}}
\vspace{-.9cm}
\caption{The same as in Fig. 10, but for a $d$ quark inside  
the proton. } 
\end{figure}

\section{CONCLUSIONS}

 A microscopical model for hadron em form
factors in both SL and TL region has been proposed. In our model
 the quark-photon vertex for the process 
where a
virtual photon materializes in a $q \bar{q}$ pair 
is approximated by a microscopic VMD model plus a bare term.

 
 Both for the pion 
and the nucleon good results are obtained in the SL region. 
  The Z-diagram (i.e. higher Fock state component) has been shown to be essential, 
in the adopted reference
frame  ($q^+ \ne 0$).
 The possible zero in $G^p_E
 \mu_p/G^p_M$ turns out to be related to the pair-production contribution.
 In the TL region our calculations give a fair description of the proton and pion data,
 although some strength is lacking for 
 $q^2 = 4.5$ ~ (GeV/c)$^2$ ~ and ~ $q^2 = 8$ ~ (GeV/c)$^2$.
 
 The analysis of nucleon form factors allows us to get
 a phenomenological Ansatz for the nucleon LF wave function,
 which reflects the asymptotic behaviour suggested by the one-gluon-exchange dominance. 
 This LF wave function is then
 used to evaluate the unpolarized transverse momentum distributions
 and the longitudinal momentum distributions of the quarks in the nucleon.
 
 Our next step will be the calculation of polarized transverse and longitudinal momentum distributions.

\end{document}